\documentclass[%
 reprint,
 amsmath,amssymb,
 aps,
]{revtex4-2}
\usepackage{xcolor}
\usepackage{graphicx}
\usepackage{dcolumn}
\usepackage{bm}

\begin{document}

\preprint{}

\title{ Competition between superconductivity and molecularization in quantum nuclear behavior of Lanthanum Hydride}

\author{S. van de Bund}
\author{G. J. Ackland}%
 \email{gjackland@ed.ac.uk}
\affiliation{Centre for Science at Extreme Conditions and School of Physics and Astronomy, University of Edinburgh, Edinburgh, U.K.}
%


\date{\today}

\begin{abstract}
Lanthanum hydride is the superconductor with the highest known critical temperature.
It is believed that the superconductivity is of standard BCS-type, with electrons forming Cooper pairs and opening the superconducting band gap.   Here we show that the BCS electron pairing is in competition with an alternative pairing: covalent bonding.
We show that the covalent pairing is favored at lower pressures, the superconducting cubic phase becomes unstable as pressure is reduced.  Previous calculations based on static relaxation neglect three factors, all of which are important in stabilizing the cubic phase.  Finite temperature plays a role, and two quantum effects are also important - the nuclear wavefunction contributes to destablizing the H$_2$ molecules, and the zero-point pressure means that calculated pressures are significantly overestimated by standard methods.    We demonstrate these phenomena using Born-Oppenheimer and path-integral molecular dynamics:  these give the same qualitative picture, with  nuclear quantum effects (NQE) increasing the transition pressure significantly.
This competition between molecularization and superconducting gap formation is the fundamental reason why hydride superconductors have so far been found only at high pressure.
\end{abstract}

\maketitle



It has long been believed that atomic, metallic hydrogen will be a high T$_c$ superconductor, based on the electron-phonon coupling (BCS) theory\cite{ashcroft1968}.  The rationale is that low-mass hydrogen lends itself to high phonon frequencies.
The practical difficulty with hydrogen superconductors is that instability against forming Cooper pairs is in competition with a more common pairing effect - molecule formation.
In many superhydrides, H$_2$ molecular electron states lie well below the Fermi energy\cite{marques2023}. 
 Meanwhile symmetry-breaking  phonon instabilities provide another way of lowering the free energy by opening pseudogaps at the Fermi energy\cite{ackland2004origin}.  Molecule formation can be suppressed by very high pressures, by chemical doping, or both.  The doping approach can be understood theoretically as adding extra electrons (from cations) to atomic metallic hydrogen to suppress molecule formation and thereby induce superconductivity\cite{Ashcroft2004a,zurek2009little,peng2017,zurek2019,laniel2022high}.

Such materials have been made in situ in diamond anvil cells by increasing pressure in a hydrogen-rich environment\cite{eremets2022high}.  The first example of this type of superconductor, hydrogen sulfide, has been well characterized experimentally with a cubic structure, and can be understood theoretically by standard density functional calculations  to have  composition H$_3$S and conventional (BCS) superconductivity\cite{Drozdov2015,errea2015high}.   

There have been a number of recent demonstrations and reports of high temperature superconductivity in hydrogen-rich compounds at high pressure: The experiments are challenging, and complementary density functional calculations (DFT) are essential to determine hydrogen composition and structure. Calculations also demonstrate that high $T_c$ can arise from conventional (BCS) superconductivity\cite{kostrzewa2020lah10,pickard2020superconducting,chen2021phase,sun2020second,hilleke2022tuning}. 
The current experimental record $T_c$ stands at 250-260 K in a Lanthanum hydride, believed to be LaH$_{10}$.\cite{drozdov2019superconductivity,somayazulu2019evidence}   
However, a number of issues around understanding this material and how to improve on it remain unresolved\cite{struzhkin2020superconductivity}.  
In particular,  
X-ray diffraction shows that the superconducting phase has  a face-centered cubic ($Fm\overline{3}m$) arrangement of lanthanum atoms\cite{drozdov2019superconductivity,somayazulu2019evidence}. 

However, density functional calculations suggest that $Fm\overline{3}m$ is  unstable to soft phonon distortions primarily involving hydrogen.  Such distortions produce infinities in the Migdal-Eliashberg equation and invalidate the BCS approach: the typical theoretical approach is then to ignore the unstable modes\cite{wang2019pressure,papaconstantopoulos2020high,kwang2020superconducting,kruglov2020superconductivity,elatresh2020optical}, or to assume the cubic structure is stable and that the unstable mode frequencies can be renormalized \cite{errea2020quantum}.
With such approximations, standard BCS calculations of $T_c$ via density functional theory and the Migdal-Eliashberg equations then suggest a similarly high value for $T_c$, agreement with experiment typically being within a factor of 2\footnote{Better agreement is often reported, but this can be traced to tweaking of undetermined parameters in the model}. 

Below 150 GPa $Fm\overline{3}m$ LaH$_3$ has imaginary phonons  
throughout the Brillouin zone, and instabilities remain at some points up to 220 GPa\cite{errea2020quantum}.   Calculations showing imaginary phonons are not unusual in real materials, and are normally linked with temperature-driven phase transitions.  Imaginary phonons arise when the curvature of the interatomic potential is negative is some direction: the high symmetry structure is a saddle point in energy E($\{r\})$.
In the Born-Oppenheimer approximation, an imaginary phonon means that the structure is unstable at 0 K.    Nevertheless, there may be several nearby energy minima and if temperature is high enough for the system to  sample all of these, then the high-symmetry position may also be the average one, as observed in experiment.   
In addition to temperature, the NQE zero-point motion to the hydrogen atoms may be sufficient to avoid being trapped in a potential minimum.  Indeed perturbative approach to zero-point energy,  the stochastic self-consistent harmonic approximation, was applied to LaH$_{10}$ and shown to eliminate all instabilities\cite{belli2021strong,errea2020quantum}.

Here, we investigate LaH$_{10}$ with DFT forces and both Born Oppenheimer (BOMD) and path integral molecular dynamics (PIMD) to determine the phase stability.  We find no qualitative difference between the two methods: both show a molecular to atomic transition with pressure.  However, NQE significantly favour the atomic phase, and make a significant contribution to the calculated pressure.


\begin{figure}
    \centering
  
        \includegraphics[width=0.49\columnwidth]{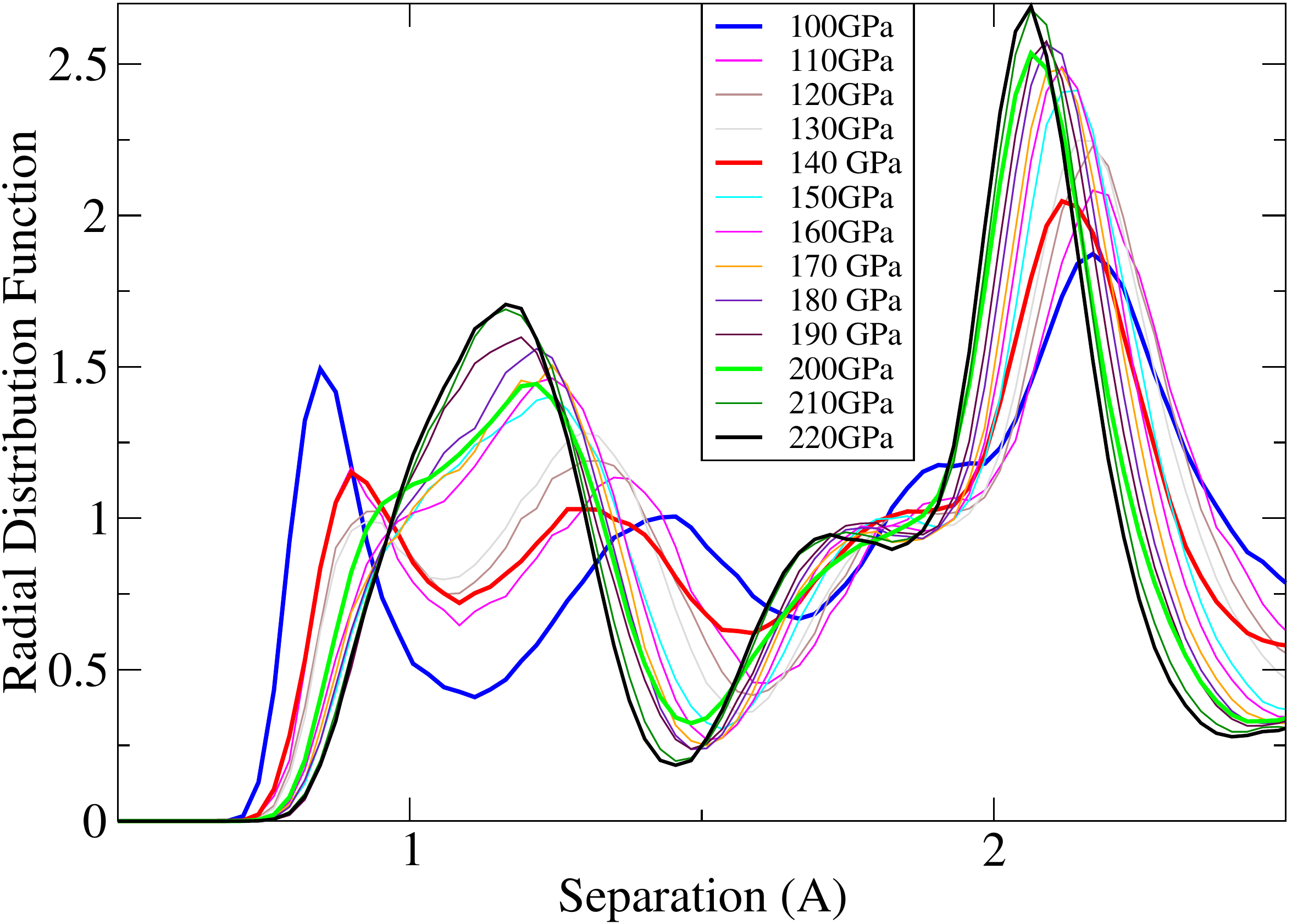}
        \includegraphics[width=0.49\columnwidth]{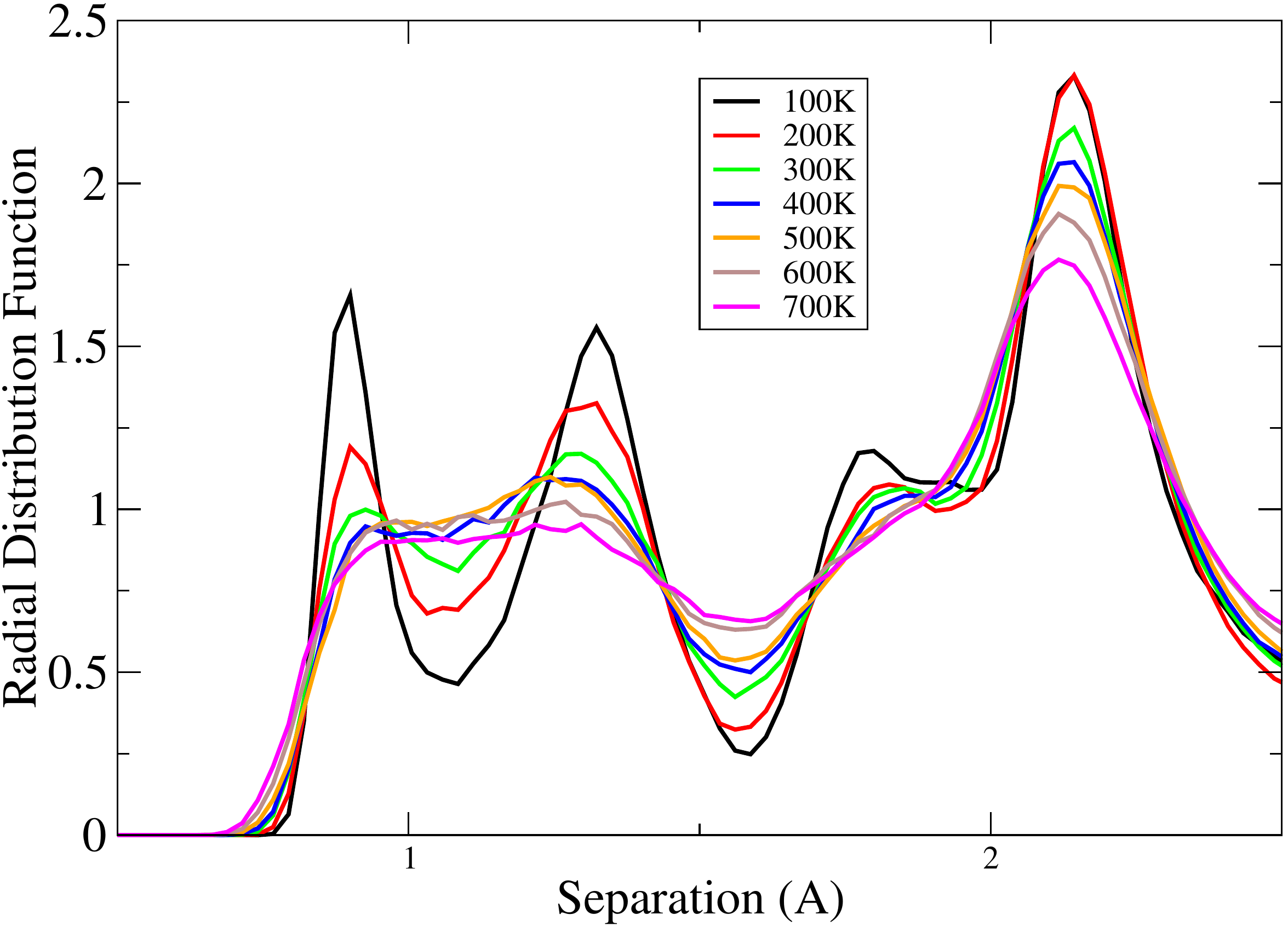}
    \includegraphics[width=0.47\columnwidth]{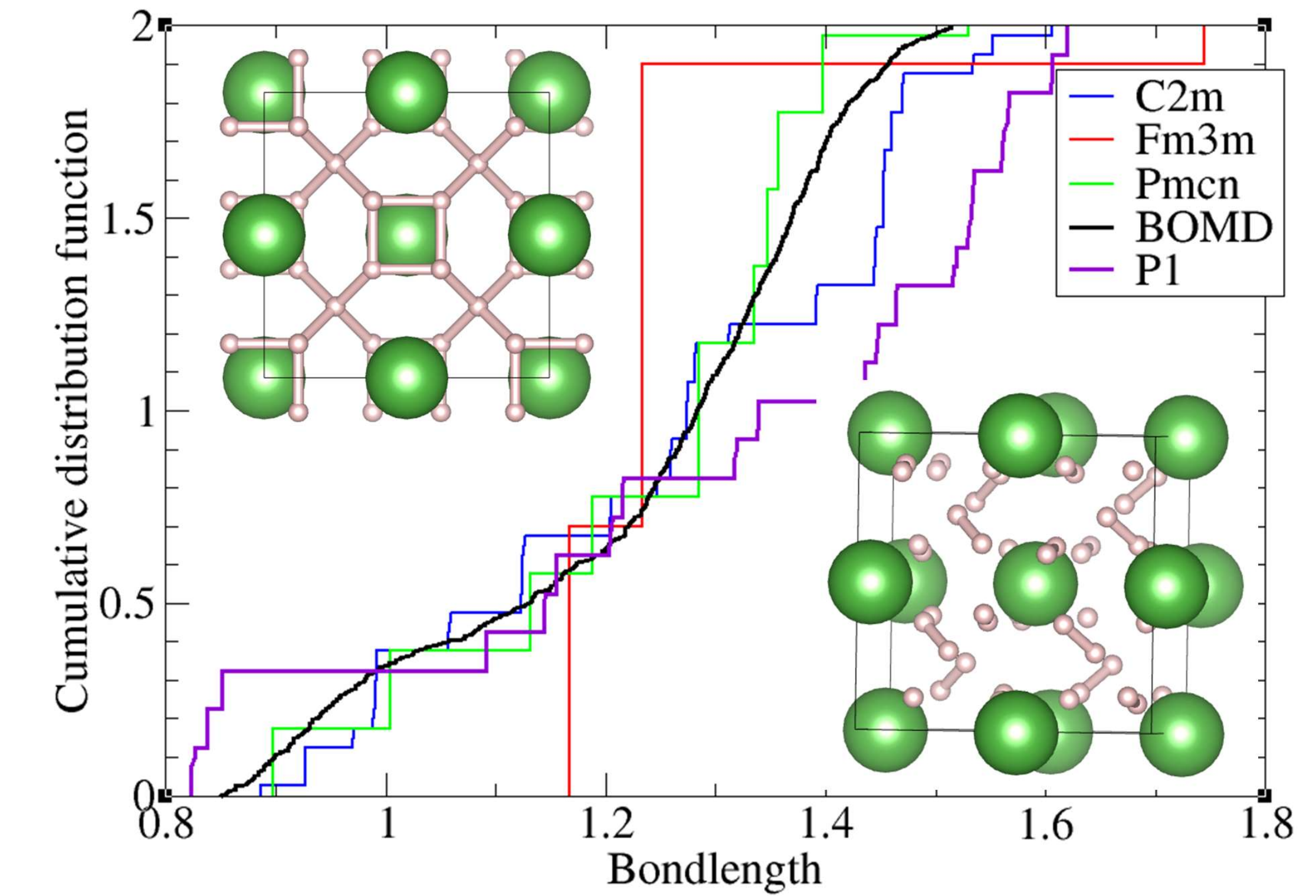}      \includegraphics[width=0.49\columnwidth]{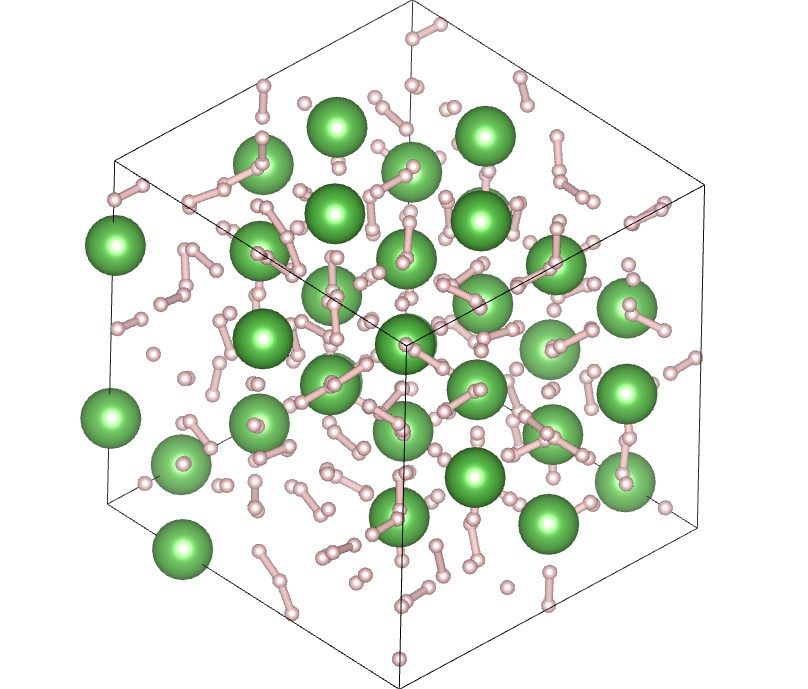}
    \caption{ (a)
    Radial distribution function from BOMD at various pressures and 300 K.
    (b)   Radial distribution function from BOMD at various temperatures and 140 GPa.  The large peak above 2 \AA is from H-La. 
    (c) Cumulative Distribution Function (HH pairs per H-atom) for the three static relaxed structures with $Fm\overline{3}m$, $Pbcn$ and $C2m$ symmetry, and one relaxation from a large BOMD snapshot.  Insets show structures at 150 GPa $Fm\overline{3}m$ (``HH bonds'' less than 1.6 \AA\, shown) and  $P1$ La$_4$H$_{12}$(H$_2)_{14}$ (``HH bonds'' less than 1.0 \AA)
    d) Relaxed snapshot from 150 GPa NPT ensemble MD run. }
    \label{fig:rdf}
\end{figure}

First, we report 
PIMD  which treats all the nuclei as quantum particles.  It is a precise method for calculating of the nuclear wavefunction.  PIMD  samples the density matrix (finite temperature nuclear wavefunction) using a set of coupled classical BOMD simulations (``beads''), connected by springs.  In the Copenhagen interpretation, a given bead represents a possible outcome of a measurement of all atomic positions.
Each hydrogen atom position is delocalized due to thermal and quantum broadening, which are represented by a combination of classical motion within a bead, and the spread of positions of the atom across different beads (radius of gyration). Even at 300 K, the radius of gyration of hydrogen - a measure of quantum broadening -  is found to be at least $\sim $0.15 {\AA}, comparable with thermal effects. 
Note that the electron-phonon coupling arises only through the connections between beads, so the explicit BCS coupling terms are absent and the condensation of Cooper pairs cannot occur.

\begin{figure}[!htb]
    \centering
    \includegraphics[width=0.95\columnwidth]{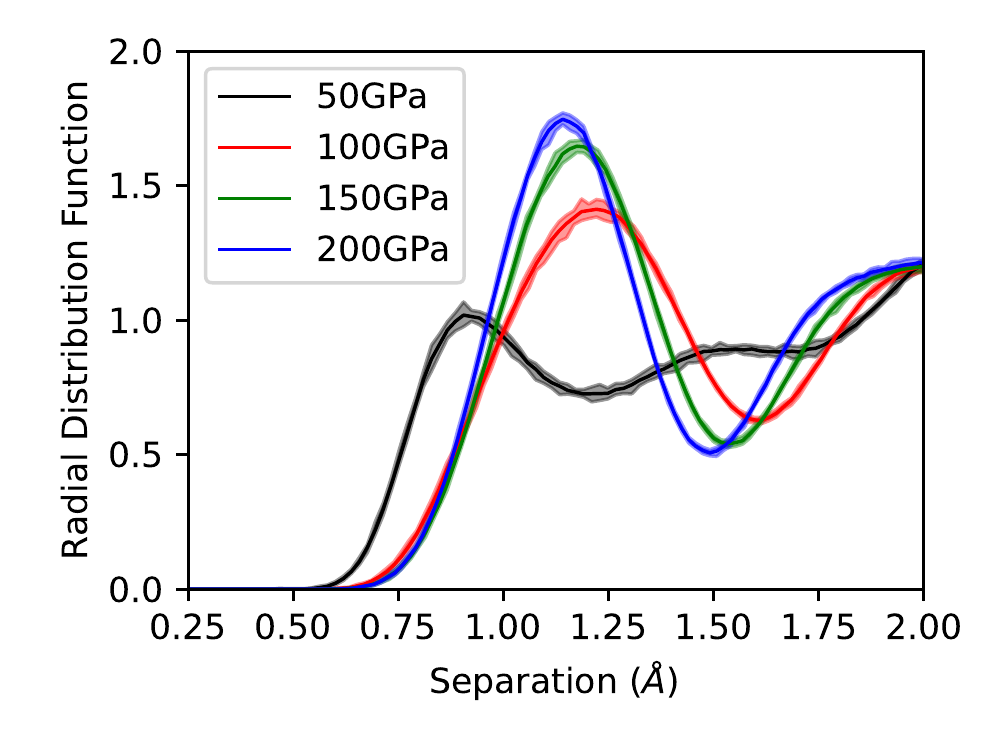}
    \caption{Radial distribution functions from PIMD at 300 K and various pressures. Filled areas indicate the minimum and maximum extent of the RDF across all beads, and can be thought of as the resulting uncertainty due to NQE.}
    \label{fig:pimd-rdf}
\end{figure}

We find that at low pressures, the PIMD shows a distinctive covalent bonding peak at 50 GPa (Fig. \ref{fig:pimd-rdf}). 
At pressures of 100 GPa and higher, the PIMD ensemble-averaged structure adopts the highly-symmetric $Fm\overline{3}m$ structure, even though transient bonding occurs within each bead. The additional NQE together with the thermal contribution is likely sufficient to overcome the imaginary modes that lead to distortions in the zero temperature structures-- the behavior is otherwise qualitatively similar to the classical nuclei BOMD, but shifted by $\sim$20 GPa.


\begin{figure}[!htb]
    \centering
    \includegraphics[width=0.9\columnwidth]{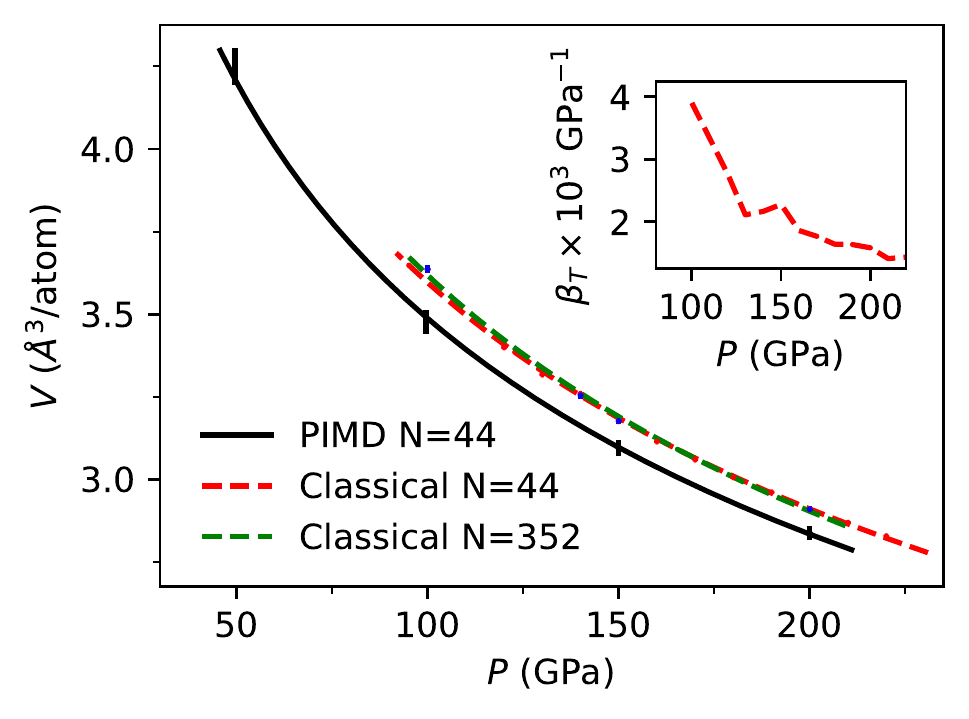}
    \caption{Comparison of equation of state between PIMD and AIMD at 300 K, showing the resulting pressure shift. Lines are obtained with a Vinet EOS fit\cite{vinet1987temperature}. For comparison, the experimental volume at 175 GPa is 3 \AA$^3$/atom\cite{somayazulu2019evidence}
    Calculations using BLYP, showing functional dependence with BLYP having a larger volume by around $0.2$ \AA$^3$/atom, are given in SM.   Inset, compressibility of BOMD calculation for 44 atoms at 300K, showing the  peak at 150 GPa corresponding to the transition from molecular to atomic.
    \label{fig:pimd-eos}}
\end{figure}

Born-Oppenheimer molecular dynamics treats all the nuclei as classical particles.  It samples the full potential surface, unlike anharmonic lattice dynamics which implicitly represents the potential as Taylor expansion about a predetermined average structure.  By sampling the phase space, BOMD correctly incorporates all classical entropy effects, and explicitly ignores all quantum effects. 
Our simulations are done using the CASTEP code\cite{CASTEP} with {\it de facto} standard PBE exchange-correlation\cite{PBE} and NPT ensemble with Nose-Hoover thermostat, cell-quench equilibration and constant-stress Parrinello-Rahman barostat. We used cubic La$_{32}$H$_{320}$ supercells at a range of pressures and temperatures.   Convergence testing shows that k-point sampling of at least 3x3x3 is required, to avoid the MD moving to structures which are tuned to the k-point set. Obtaining quantitative accuracy for the atomic-molecular transition in hydrogen is particularly challenging for DFT, because the exchange-correlation function must give  a good description of both atomic and covalent hydrogen. PBE tends to form overly weak H$_2$ bonds, and to test sensitivity to exchange-correlation functional we repeated some calculations with the BLYP functional\cite{blyp-B,blyp-LYP}, which has the opposite tendency\cite{clay2014benchmarking,azadi2017role}. 
Simulations were run as static relaxations and at 300 K and pressures between 50 and 220 GPa.

These large simulations are augmented by smaller 44-atom BOMD, the same size as used in PIMD: these show good agreement with their larger counterparts, indicating that the PIMD calculations are also converged with system size. The PIMD runs were performed using the i-Pi code\cite{Kapil2019i-PI} together with Quantum Espresso\cite{QE-2017}. 16 beads were used for all simulations, in a constant stress ensemble with a stochastic PILE\_L thermostat\cite{Ceriotti2010Efficient} and anisotropic stochastic barostat\cite{Ceriotti2014i-PI} as implemented in i-PI, with identical k-point sampling grid as in the BOMD runs.


The mean-squared displacement of atoms (which plateaus at a low value), the variation of lattice parameters (which remain broadly unchanged) and radial distributions functions (Shown in Figure \ref{fig:rdf}).  By contrast the radial distribution function shows a strong change in the bonding.  At lower pressures, e.g. 100 GPa there is a distinctive peak at 0.85 \AA, characteristic of covalently-bonded molecular hydrogen.  This is not simply an anharmonic phonon distortion from $Fm\overline{3}m$ - it represents a complete change in the nature of the bonding. As pressure is increased, the bond gets longer.  At 150 GPa, the ``bond'' peak is merely a shoulder.

BOMD can also find phase transitions, although for LaH$_{10}$ the high La mass makes exploring that phase space impractical.  We use BOMD with fictitious masses (M$_\text{La}=$2M$_\text{H}$) and NPT ensemble to better sample the equilibrium phase space at the expense of correct dynamics.  We found that at all pressures structures with the La atoms close to $Fm\overline{3}m$ remain stable, however, the transition from molecular to atomic hydrogen manifests as a distinct peak in the compressibility, indicating a two-state system rather than a continuous weakening of bonds\cite{ackland2021two}.

Static relaxations of snapshots show structures with well defined molecules.  Fig.\ref{fig:rdf}b illustrates the salient features of these static structures by plotting the number of HH pairs in the cell per hydrogen, as a function of the pair separation.  In these units the octet-rule compound LaH$_3$(H$_2)_{3.5}$ would have a CDF of 0.35 at the hydrogen bondlength. The lowest energy structure we found from 
44 atom MD with BLYP has precisely this, with 14 well-defined H$_2$ units and La remaining close to fcc positions (initially $P1$, relaxing to $P2_1$/$c$).
Considering previously reported structures, all are less stable, and break the octet rule. In $Fm\overline{3}m$ there are no short HH distances, then a large group appears around 1.2 \AA.  At 150 GPa  a lower enthalpy is obtained  by distorting in a way to create molecules with H-H separation.  $Pmcn$ symmetry generates 4 equivalent sites, so even with a quadrupled unit cell a fractional number of objects (e.g. 3.5 H$_2$'s) is forbidden. In fact, the 44 atom $C2m$ structures forms 8 covalent-style bonds (0.898 \AA, 0.95 Mulliken population)  and 8 more H$_2$ pairs (1.004 \AA, 0.70 e), arguably LaH$_2$(H$_2)_4$.   Further symmetry breaking to $C2m$\cite{geballe2018synthesis} finds another solution to the incompatibility between charge and stoichiometry, H$_3^-$ units:  La$^{3+}_4$H$^-_{10}$(H${_3^-})_2$(H$_2)_{12}$. 

An important feature of the BLYP calculation is that the density is significantly lower than for PBE, by about 7\%, similar in magnitude to the quantum nuclear effects.  This equates to about 40GPa and gives some indication of the uncertainty in DFT calculations.


Thus the classical DFT situation reflects conventional chemistry, with the LaH$_{10}$ stoichiometry demanding large unit cells to satisfy the octet rule with La$^{3+}$, H$^-$ and H$_2$.  Relaxations from 352 atom MD simulation gave similar patterns: molecule formation with no obvious symmetry and would induce small distortions from cubic which may be detectable \cite{geballe2018synthesis,laniel2022high}.    $Fm\overline{3}m$ has special symmetry positions which indicate a specific LaH$_{10}$ stoichiometry, but the low symmetry of the molecular phase has no such constraints, so off-stoichiometry compounds on the type LaH$_3($H$_2)_x$ are expected.  Indeed, no fewer than seven different stoichiometries were recently synthesized at pressures below the atomic-molecular transition\cite{laniel2022high}: reassuringly, none of them require $x<0$.

The quantum effects in LaH$_{10}$ can be separated into two parts, the effect on pressure and the effect on structure.  Fig \ref{fig:pimd-eos} shows that for a given volume, the pressure calculated by PIMD is typically 20 GPa lower than for classical simulation.   This means that all previous DFT calculations have overstated the pressure to which their results apply.  With respect to comparison with experiment, it means that the ``150 GPa'' transition from molecular to atomic with standard DFT corresponds to 130 GPa, below where the structure has been reported experimentally\cite{drozdov2019superconductivity}.  

In summary, we have shown that  
including quantum nuclear effects in DFT calculations on LaH${12}$ does not have a quantitative effect, rather it shifts the calculated pressure of transformation from molecular/metallic to atomic/superconducting structure to lower pressures. Increasing temperature has a similar effect, of similar magnitude, and has the advantage over switching off quantum mechanics of being experimentally possible.   There is also considerable dependence on exchange-correlation functional leading to uncertainty in the calculation 

Perhaps our most wide-ranging result is that the BCS superconducting phase of LaH$_{10}$ is in competition with molecularization, a feature which appears to be true in many high-pressure hydrides\cite{marques2023}.  Molecular phases have lower energy, but also lower density and entropy: this explains why the high-T superconductors found to date require high pressure.  

\bibliography{apssamp}

\end{document}


\title{\textbf{Supplementary materials for ``Competition between superconductivity and molecularization in quantum nuclear behavior of Lanthanum Hydride''}}

\author{S. van de Bund,$^1$ \& G. J. Ackland$^1$}
\maketitle

\begin{enumerate}
    \item \small Centre for Science at Extreme Conditions and School of Physics and Astronomy, University of Edinburgh, Edinburgh, U.K.
\end{enumerate}

\section*{Electronic Density of States}
\begin{figure}[!htpb]
    \centering
    \includegraphics[scale=0.75]{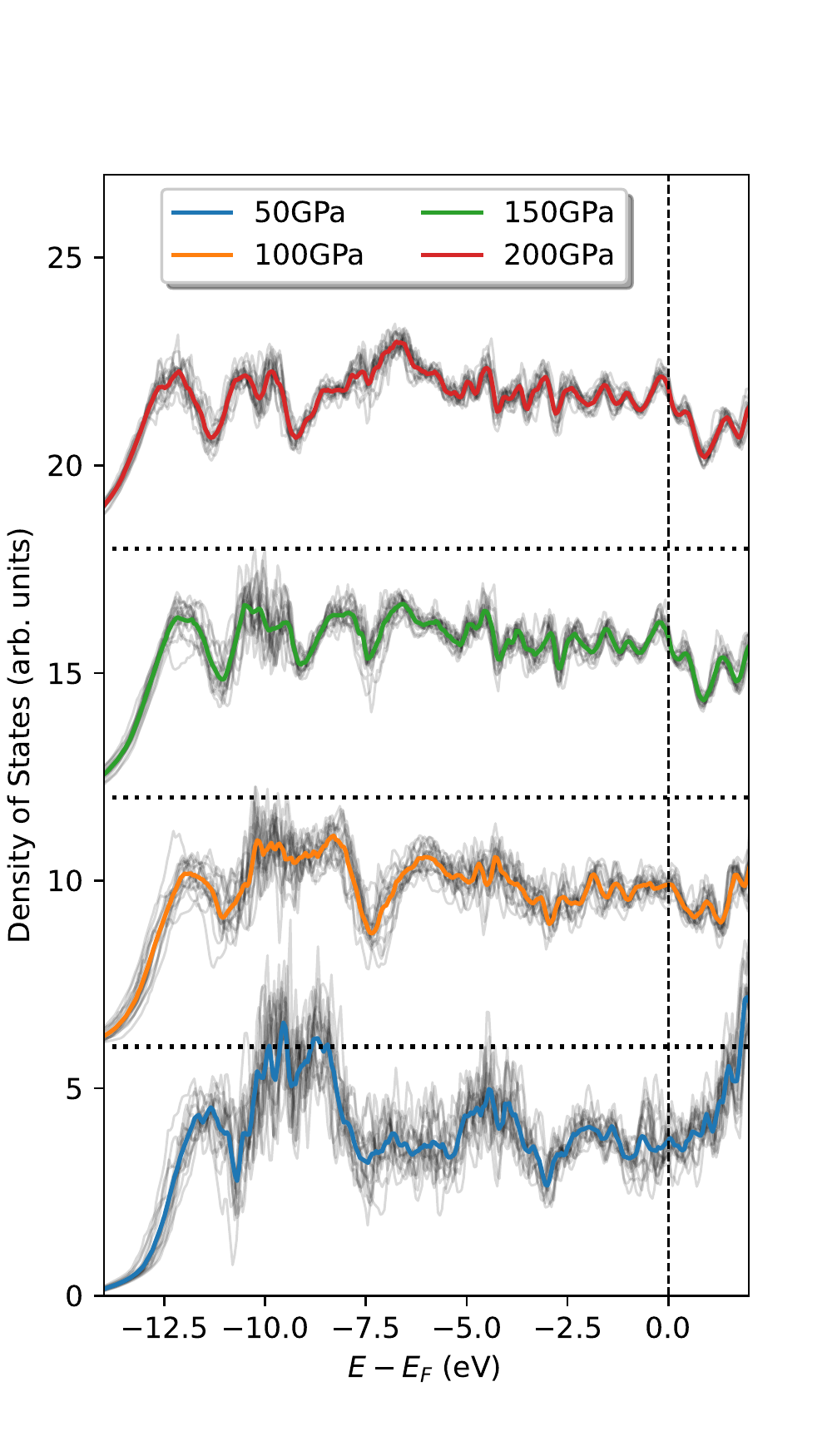}
    \caption{Electronic DOS for the PIMD simulations. The DOS corresponding to each bead is shown in the semi-transparent background lines, with the overall averaged DOS overlaid.}
    \label{fig:pimd-dos}
\end{figure}

The electronic density of states were calculated using an off-centered 16x16x16 k point grid in Quantum Espresso. The results are shown in Fig. \ref{fig:pimd-dos}.

Averaging was done by selecting 10 equally spaced snapshots for each of the 16 beads trajectories.
Because each bead contains a different nuclear configuration, the corresponding electronic DOS varies quite significantly across the beads, most notably so in the low pressure case. The main effect of the averaging is to smooth out the features of the DOS. Across the transition, there is a slight increase in the value of the DOS at the Fermi energy, but otherwise the DOS remains mostly flat, likely due to the smoothing from the PIMD.

\section*{APPENDIX: Convergence testing}
We tested finite size effects, plane wave cutoff and k-point sampling for the molecular dynamics.   Surprisingly, even systems as small as 44 atoms showed the same change from atomic to molecular with pressure.  The main calculations were done at the CASTEP default cutoff 463 eV: repeating the calculation at 408 eV gave indistinguishable results.

Larger system size, lower temperature and small k-point sampling tend to favour molecular formation.  The bold lines in the figure show three MD runs with 356 atoms.  The long-range structure is remarkably consistent between all three, representing the crystal structure.  By contrast, the first peak ``molecular'' peak is clearly defined at 300 K, but not at 600 K.  Allowing the lanthanums to move more easily by setting their mass to 2 gives the same structure as with full-mas lanthanum, with slightly better definition of the molecular peak.  The thinner lines show various calculations with 44 atoms cells, showing that high k-point sampling densities facilitate the bond breaking.  All k-point sampling was done using offset Monkhorst Pack grids.

\begin{figure}[h]
    \centering
        \includegraphics[width=0.51\columnwidth]{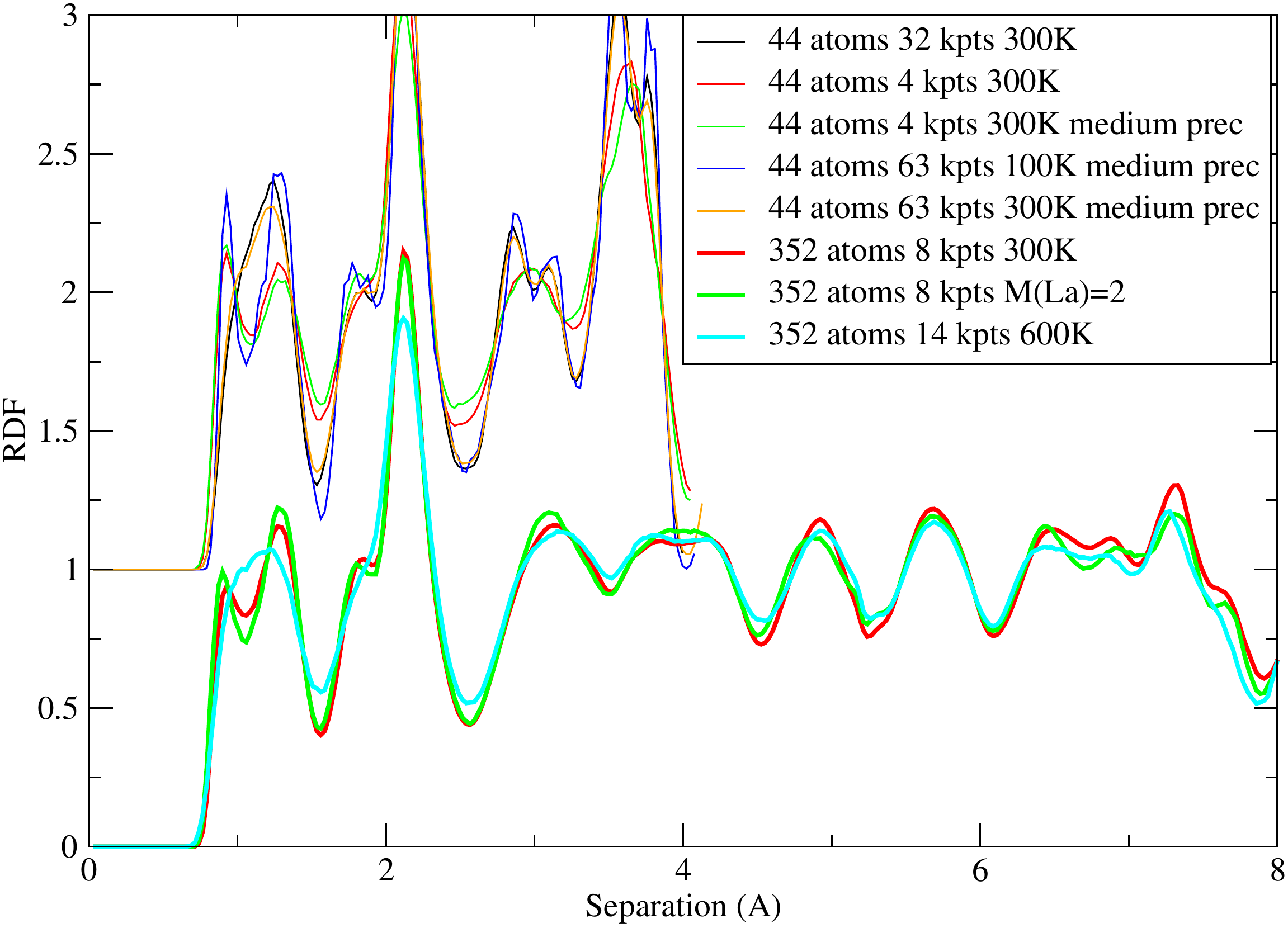}
    \caption{SUPPLEMENTAL 
    Radial distribution function from molecular dynamics at 150 GPa, at various temperature and simulation settings.  Regular calculations use a 463 eV cutoff, ``medium precision'' uses 408 eV.   }
    \label{appfig:rdf}
\end{figure}

\clearpage

\section*{APPENDIX: Functional sensitivity}

\begin{figure}
    \centering
        \includegraphics[width=0.51\columnwidth]{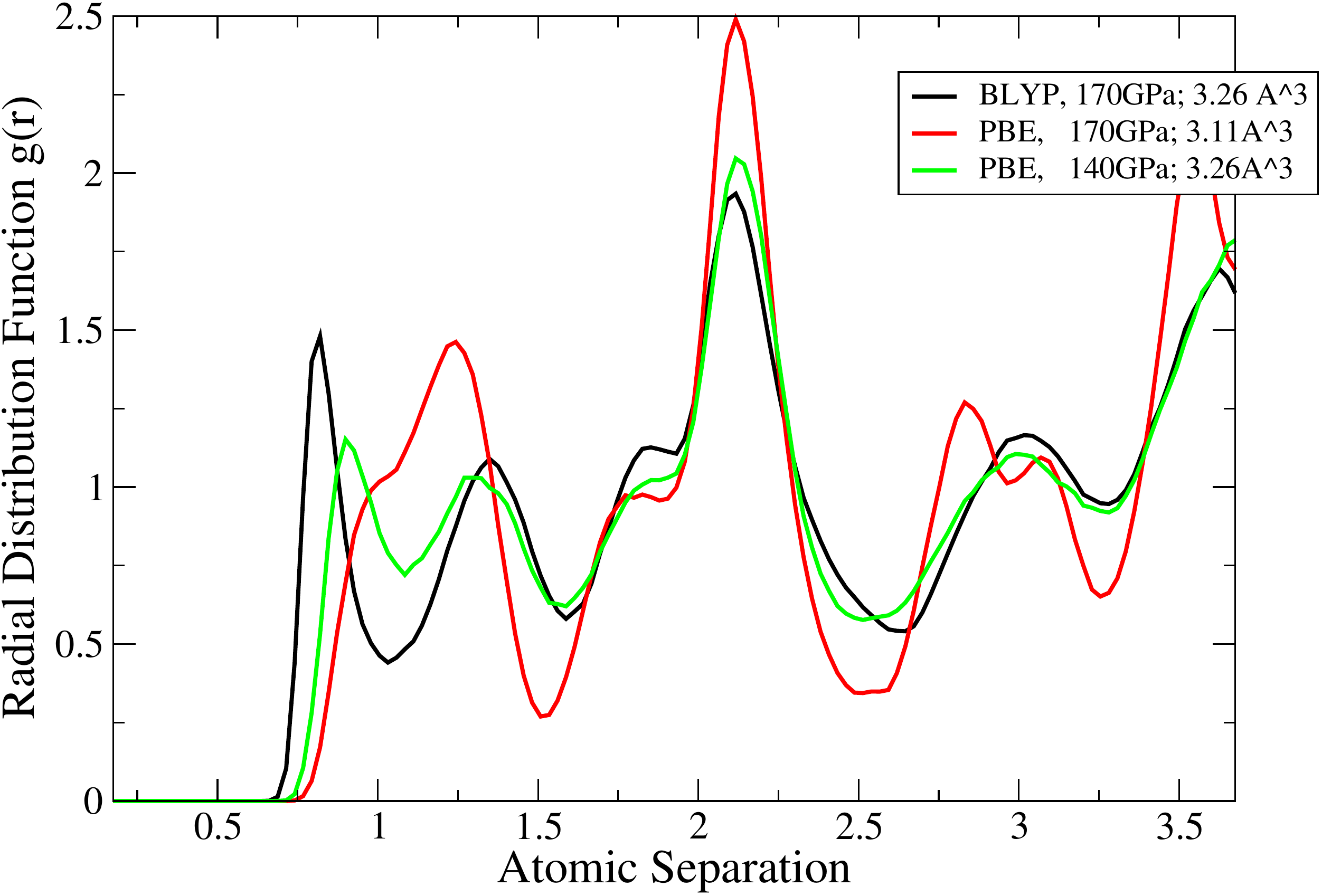}
    \caption{SUPPLEMENTAL 
    Radial distribution function from NPT molecular dynamics at 300K  comparing results with BLYP functional to PBE at the same pressure and the same density.  Even in the absence of quantum nuclear effects, PBE has  gone through the molecular-atomic transition at 150 GPa, whereas BLYP retains a strong molecular character beyond 170 GPa. 
    The large peak around 2.2 \AA$ $ is the La-H separation. }
    \label{appfig:blyp}
\end{figure}

Most calculations on high pressure hydrides are carried out using the PBE functional.   Other functionals with similar theoretical credibility, in particular those with the correct treatment of the high charge gradient limit, give stronger molecular bonding.  BLYP is an example of the latter, and in classical MD at 170 GPa at 300 K we find that BLYP gives the molecular structure while PBE is already atomic (Fig. \ref{appfig:blyp}).  Part of this may be attributed to density: PBE 3.11 \AA$^3$/atom compared with BLYP 3.26 \AA$^3$/atom, but comparing at the same density still shows considerably more molecular character with BLYP.

\begin{figure}
    \centering
        \includegraphics[width=0.51\columnwidth]{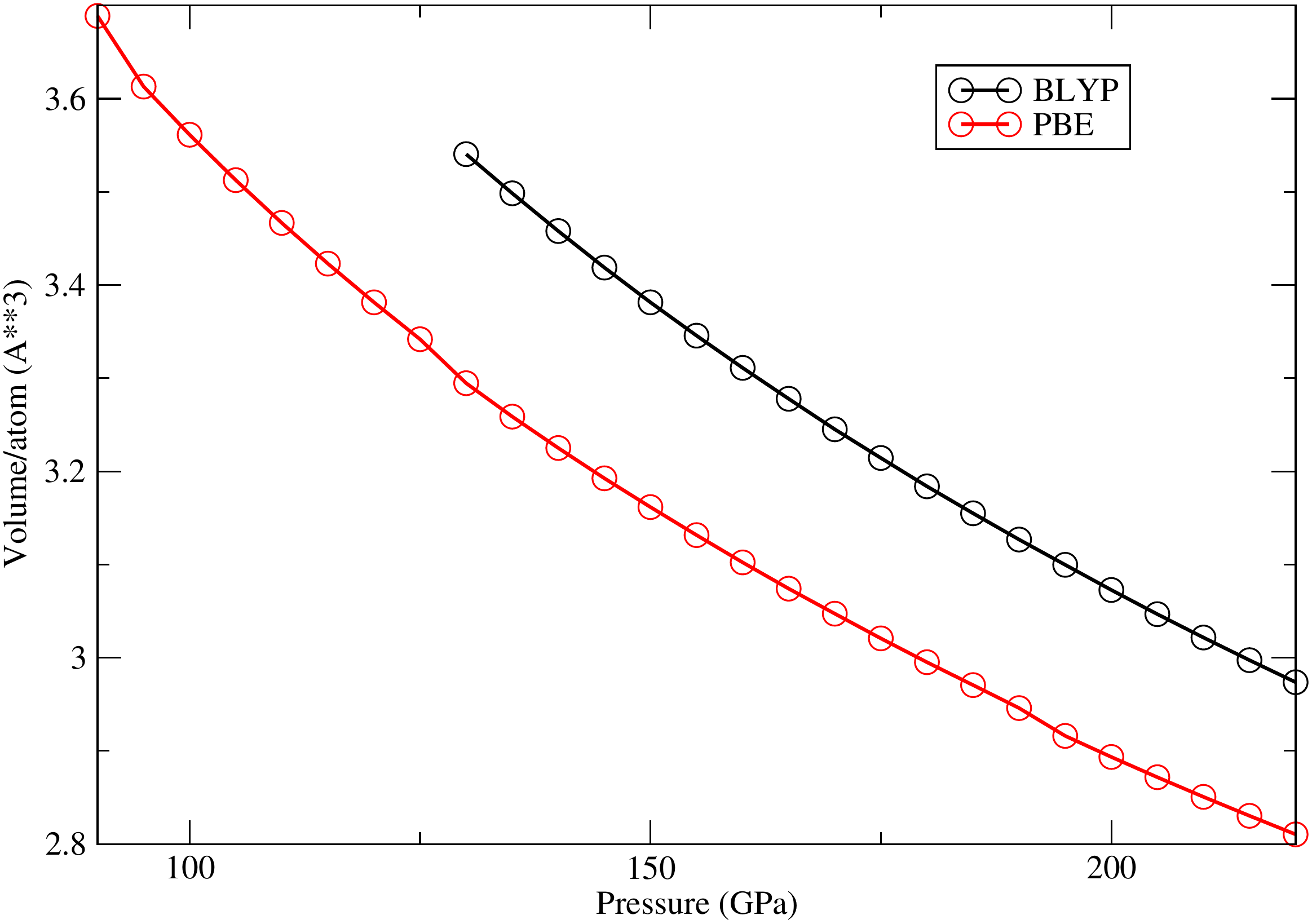}
    \caption{SUPPLEMENTAL Static relaxation of 44-atom MD snapshot calculated with BLYP and PBE functionals.
     }
    \label{appfig:XC}
\end{figure}

\begin{table}[hbt]
    \centering
    \begin{tabular}{lcccccc}
    \hline
  H &                 0.884521& 0.351747& 0.114175 \\
      H &             0.384521& 0.648253& 0.385825 \\
      H &             0.615479& 0.148253& 0.885825 \\
      H &             0.115479& 0.851747& 0.614175 \\
      H &              0.329628& 0.331721& 0.706159 \\
      H &              0.829628& 0.668279& 0.793841 \\
      H &              0.170372& 0.168279& 0.293841 \\
      H &             0.670372& 0.831721& 0.206159 \\
      H &              0.335094& 0.855630& 0.129314 \\
      H &              0.835094& 0.144370& 0.370686 \\
      H &              0.164906& 0.644370& 0.870686 \\
      H &              0.664906& 0.355630& 0.629314 \\
      H &              0.206227& 0.296735& 0.050432 \\
      H &              0.706227& 0.703265& 0.449568 \\
      H &              0.293773& 0.203265& 0.949568 \\
      H &              0.793773& 0.796735& 0.550432 \\
      H &              0.213171& 0.791308& 0.710236 \\
      H &              0.713171& 0.208692& 0.789764 \\
      H &              0.286829& 0.708692& 0.289764 \\
      H &              0.786829& 0.291308& 0.210236 \\
      H &              0.982731& 0.612122& 0.148971 \\
      H &              0.482731& 0.387878& 0.351029 \\
      H &              0.017269& 0.112122& 0.648971 \\
      H &              0.517269& 0.887878& 0.851029 \\
      H &             0.458831& 0.016431& 0.098524 \\
      H &              0.958831& 0.983569& 0.401476 \\
      H &              0.041169& 0.483569& 0.901476 \\
      H &             0.541169& 0.516431& 0.598524 \\
      H &              0.565094& 0.135965& 0.131001 \\
      H &              0.065094& 0.864035& 0.368999 \\
      H &             0.934906& 0.364035& 0.868999 \\
      H &              0.434906& 0.635965& 0.631001 \\
      H &              0.649576& 0.515304& 0.359917 \\
      H &              0.149576& 0.484696& 0.140083 \\
      H &              0.850424& 0.984696& 0.640083 \\
      H &              0.350424& 0.015304& 0.859917 \\
      H &              0.847280& 0.687696& 0.209357 \\
      H &              0.347280& 0.312304& 0.290643 \\
      H &             0.652720& 0.812304& 0.790643 \\
      H &              0.152720& 0.187696& 0.709357 \\
            La    &    0.026089& 0.471571& 0.509777\\ 
            La    &     0.526089& 0.528429 & -0.009777 \\
            La     &    0.473911& 0.028429& 0.490223 \\
            La      &     -0.026089 & -0.028429& 0.009777\\

\hline
\end{tabular}
\caption{Crystal structure (fractional coordinates) for P2$_1$/c, a low energy molecular structure from BLYP BOMD at 215 GPa. Lattice parameters:\quad\quad\quad\quad
        a =      4.797 \AA, \, 
        b =      5.386 \AA, \,
        c =      5.140 \AA, \,
        $\alpha$ =   96.78$^\circ$, \,
        $\beta$  =   90.00$^\circ$, \,
        $\gamma$ =   90.00$^\circ$.
}
\end{table}
